\documentclass[prl,aps,amssymb,twocolumn]{revtex4-1}
\usepackage{amsmath}
\usepackage{amssymb}
\usepackage{amsthm}
\usepackage{amsfonts}
\usepackage{listings}
\usepackage{enumerate}
\usepackage{latexsym}
\usepackage{color}
\usepackage[colorlinks=true,urlcolor=blue,citecolor=blue,linkcolor=blue]{hyperref}
\usepackage{psfrag}

\usepackage{bm}
\usepackage{graphicx}
\usepackage{subfigure}

\newcommand{\beq}{\begin{equation}}
\newcommand{\eneq}{\end{equation}}

\newcommand{\bal}{\begin{align}}
\newcommand{\eal}{\end{align}}

\input{epsf}

\begin{document}

\tolerance 10000

\newcommand{\vk}{{\bf k}}

\title{Time-Reversal-Breaking Weyl Fermions in Magnetic Heusler Alloys}

\author{Zhijun Wang$^{1}$}
\author{M. G. Vergniory$^{2,3}$}
\thanks{Zhijun Wang and M. G. Vergniory contributed equally to this work}
\author{S. Kushwaha$^{4}$}
\author{Max Hirschberger$^{1}$}
\author{E. V. Chulkov$^{2,5,7}$}
\author{A. Ernst$^{6}$}
\author{N. P. Ong$^{1}$}
\author{Robert J. Cava$^4$}
\author{B. Andrei Bernevig$^{1}$}
\affiliation{${^1}$Department of Physics, Princeton University, Princeton, NJ 08544, USA}
\affiliation{${^2}$Donostia International Physics Center, P. Manuel de Lardizabal 4, 20018 Donostia-San Sebasti\'an, Spain}
\affiliation{${^3}$ Department of Applied Physics II, Faculty of Science and Technology, University of the Basque Country UPV/EHU, Apartado 644, 48080 Bilbao, Spain}
\affiliation{${^4}$Department of Chemistry, Princeton University, Princeton, NJ 08540, USA}
\affiliation{${^5}$ Departamento de F\'isica de Materiales, Universidad del Pa\'is Vasco/Euskal Herriko Unibertsitatea UPV/EHU, 20080 Donostia-San Sebasti\'an, Spain}
\affiliation{${^6}$ Max-Planck-Institut f\"ur Mikrostrukturphysik, Weinberg 2, D-06120 Halle, Germany }
\affiliation{${^7}$ Saint Petersburg State University, 198504 Saint Petersburg, Russia }

\date{\today}

\begin{abstract}
Weyl fermions have recently been observed in several time-reversal invariant semimetals and photonics materials with broken inversion symmetry. These systems are expected to have exotic transport properties such as the chiral anomaly. However, most discovered Weyl materials possess a substantial number of Weyl nodes close to the Fermi level that give rise to complicated transport properties. Here we predict for the first time a new family of Weyl systems defined by broken time reversal symmetry, namely, Co-based magnetic Heusler materials XCo$_2$Z (X = IVB or VB; Z = IVA or IIIA). 
To search Weyls in the centrosymmetric magnetic systems, we recall an easy and practical inversion invariant, which has been calculated to be $-1$, guaranteeing the existence of an odd number of pairs of Weyls.
These materials exhibit, when alloyed, only two Weyl nodes at the Fermi level - the minimum number possible in a condensed matter system. The Weyl nodes are protected by the rotational symmetry along the magnetic axis and separated by a large distance (of order 2$\pi$) in the Brillouin zone.  The corresponding Fermi arcs have been calculated as well. This discovery provides a realistic and promising platform for manipulating and studying the magnetic Weyl physics in experiments.
\end{abstract}

\maketitle


Weyl fermions are theorized to exist in the standard model above the symmetry breaking scale - at low-energy scales they invariably acquire mass. As such, the search for Weyl fermions has shifted to condensed matter systems, where they appear as unremovable crossings in electronic Bloch bands, close to the Fermi level, in $3$-spatial dimensions.  They recently became a reality with the experimental discovery~\cite{PhysRevX.5.031013,Xu07082015} of the theoretically predicted ~\cite{PhysRevX.5.011029,NatCommHuang} Weyl semimetals (WSM) in the TaAs family of compounds. Topological Weyl metals are responsible for an array of exotic spectroscopic and transport phenomena such as the surface disconnected Fermi arcs, chiral anomaly and anomalous Hall effects~\cite{Nielsen1983389,PhysRevB.83.205101,Xiong413,PhysRevX.5.031023,PhysRevB.90.155316,PhysRevLett.114.206401,PhysRevB.91.165105,Burkov2014}. 

Transport in Weyl materials reveal the fundamental nature of the Berry phase in magneto transport. The chiral anomaly is reflected in the negative longitudinal magnetoresistance (NMR), which on its own stems from the Weyl fermions. A minumum number (2) of Weyls could make it easier to observe the NMR than for cases with a large number of Weyls.
Due to the fermion doubling problem, Weyl nodes appear in multiples of $2$. With time-reversal invariance, this number raises to multiples of $4$. The TaAs family of compounds exhibits $24$ Weyl nodes, due to several other mirror symmetries. This large number of Weyls gives rise to complicated spectroscopic and transport properties. By comparison, the hypothetical hydrogen atom of Weyl semimetals is a material with only $2$ Weyl nodes at the Fermi level ($E_{\mathrm{F}}$), preferably cleanly separated in momentum space and in energy from other bands. 

In previous work~\cite{xu2011}, HgCr$_2$Se$_4$ has been predicted to be a Chern ({\em not} Weyl) semimetal with two crossing points, each of which possesses chirality of 2. Also, the existence of only two Weyl nodes is also expected in the Topological/normal-insulator heterostructures with magnetic doping~\cite{burkov2011,liuweyl2014}. Unfortunately, none of these systems have been verified so far.
In this paper we predict for the first time a series of magnetic Heusler compounds that can exhibit, when alloyed, two Weyl points close to Fermi energy and largely separated in momentum space. Magnetic Heusler compounds have several advantages over the other compounds where Weyl fermions have been proposed and detected. First they are ferromagnetic half-metallic compounds with Curie temperatures up to the room temperature~\cite{nbco2sn1996}, meaning they can be of great use for spin manipulation and spintronics applications. Second, they exhibit Weyl fermions, with the large associated Berry phase of their Fermi surfaces. As such, the anomalous Hall effect and spin Hall effect in these materials is theoretically expected to be large, which has already been confirmed in initial experimental studies~\cite{Claudia2005,Stuart2013,Co2ZrSnAHE}. Third, the magnetism in these materials is ``soft," meaning that an applied magnetic field can easily change the magnetic moment direction: a similar phenomenon has been observed in GdPtBi~\cite{Claudia2016,Max2016}. Since, as pointed out in the Supplementary Material~\cite{ClaudiaMn}, the structure and direction of Weyls depends on the magnetic field, the magnetic Heuslers provide us a realistic and promising platform for manipulating and studying the magnetic Weyl physics in experiments.

Based on the first principle calculations with the magnetism oriented in the $[110]$ direction, we found two Weyl nodes, related by symmetry, appear on the same axis. Their energy relative to the Fermi level can be tuned by alloying, and we provide an estimation of the appropriate concentration of the dopant necessary to tune the Weyls to the Fermi level. We carry out a symmetry analysis and solidify our ab-initio claims of the existence of Weyl nodes by showing that they are formed by bands of opposite $C_2$ eigenvalues. Furthermore we link the existence of Weyl nodes to the existence of an inversion invariant discovered earlier\cite{PhysRevB.83.245132} in the theory of inversion symmetric topological insulators. We then obtain the structure of the Fermi arcs on 001-surface. We expand upon the results in the Supplementary Material, where we also analyze the possible topological scenarios under different potential magnetization directions, and present magnetization data on one of the synthesized materials from our proposals.

\begin{figure}
\centering
 \includegraphics[width=0.48\columnwidth]{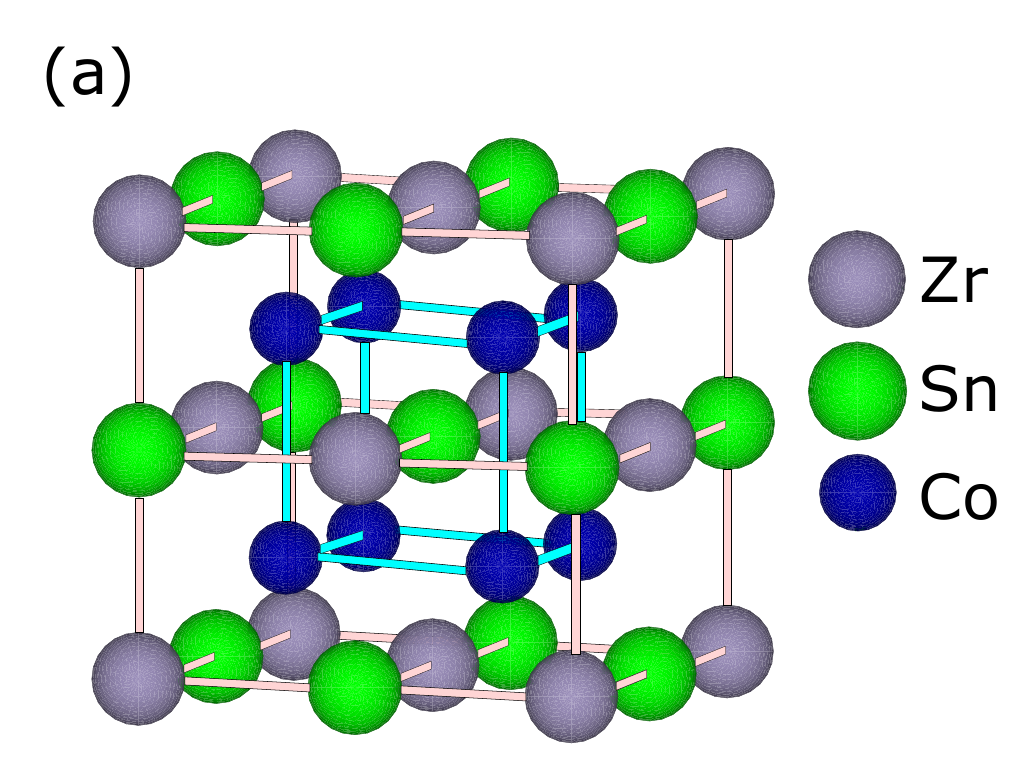}
 \includegraphics[width=0.48\columnwidth]{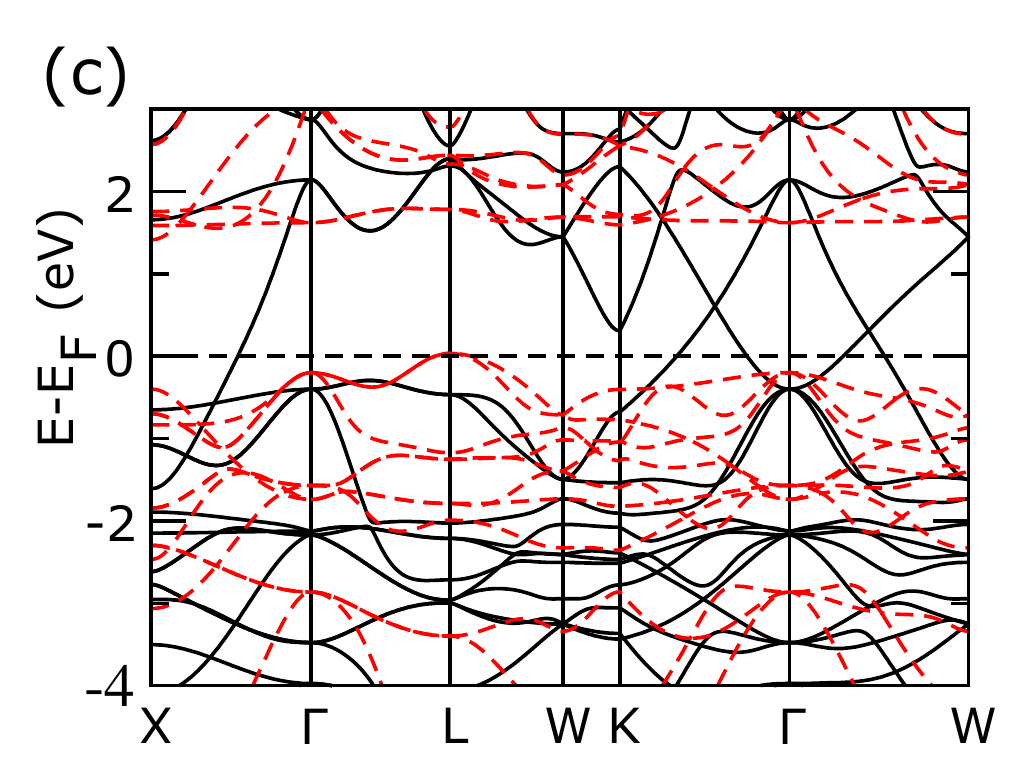}
 \includegraphics[width=0.48\columnwidth]{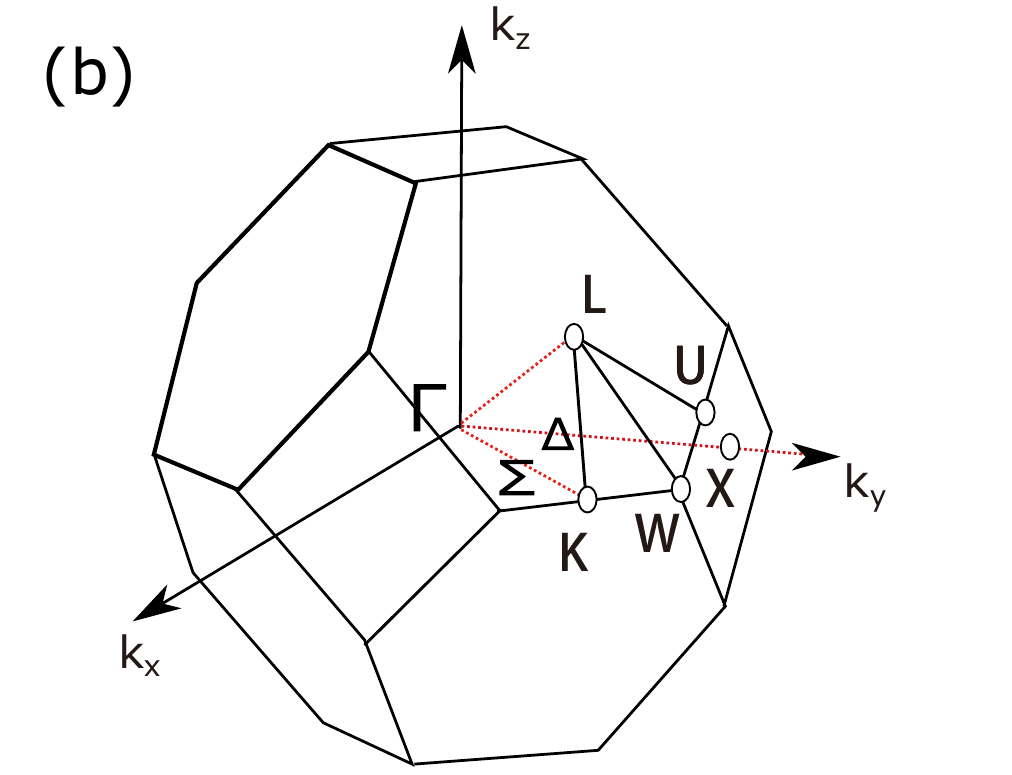}
 \includegraphics[width=0.48\columnwidth]{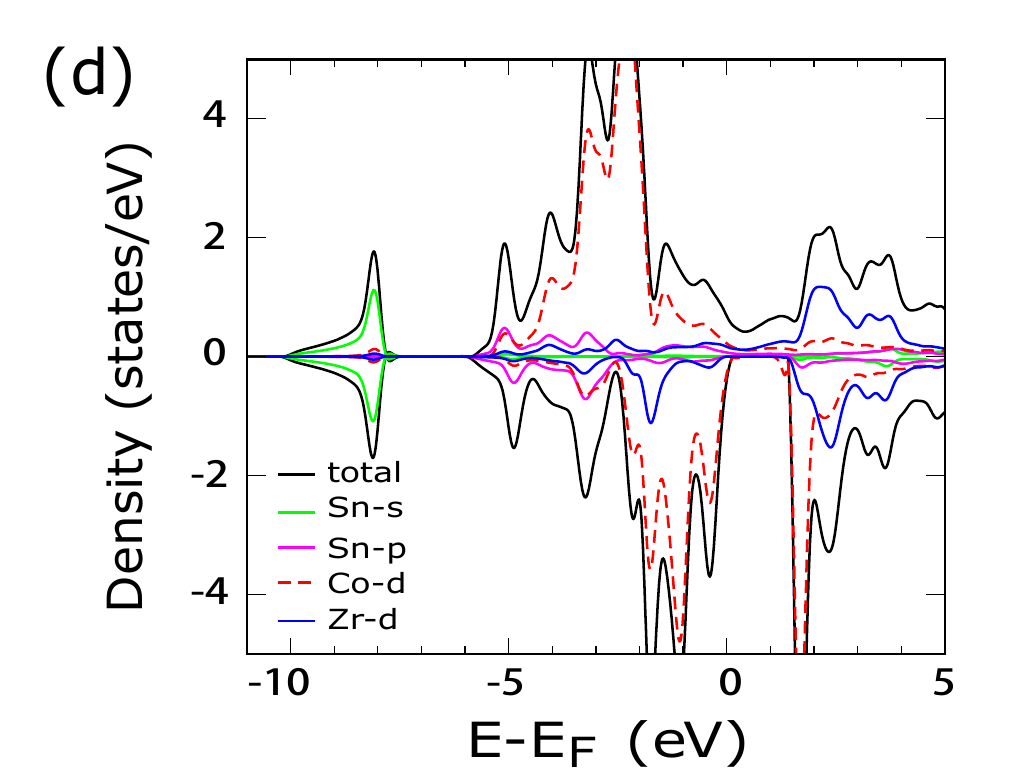}
    \caption{(a) Rocksalt crystal structure of ZrCo$_2$Sn with $Fm\bar{3}m$ space group. Co, Zr and Sn atoms are shown in blue, gray and green, respectively. (b) Brillouin zone (BZ) of the rocksalt structure.  It has three independent time-reversal-invariant points, $\Gamma$(0,0,0), $L$($\frac{\pi}{a}$,$\frac{\pi}{a}$,$\frac{\pi}{a}$), and $X$($\frac{2\pi}{a}$,0,0). (c) Bulk band structure of ZrCo$_2$Sn along high symmetry lines without spin-orbit coupling (SOC). The majority and minority spin bands are indicated by solid-black and dashed-red lines, respectively. (d) Total and partial density of states without SOC, the positive (negative) lines represent the majority (minority) spin channel.} 
     \label{fig1}
\end{figure}

Our proposed candidates for Weyl metals are the Co-based Heusler compounds: XCo$_2$Z (X = IVB or VB; Z = IVA or IIIA) with $N_v=26$, denoting the number of valence electrons ($s,d$ electrons for the transition metals and $s,p$ electrons for the main group element). This family of Co-based Heusler compounds follows the Slater-Pauling rule which predicts a total spin magnetic moment $m=N_v-24$ (the number of atoms $\times$ 6)~\cite{PhysRevB.87.024420,PhysRevB.66.174429,Felser2006}. In this paper we focus on a representative candidate -  ZrCo$_2$Sn that we have synthesized experimentally~\cite{nbco2sn1996}. We present the results for the rest of the compounds in the Supplementary Material.
We show that the interesting Weyl nodes of all the 26-electron compounds are about 0.6 eV above the Fermi energy. However, there is another Co-based Heusler family with 27-electron, that are also candidates for Weyl metals and have been synthesized experimentally~\cite{nbco2sn1996}, such as NbCo$_2$Sn and VCo$_2$Sn, making it possible to move the Weyl nodes close to or at the Fermi level by alloying. 
The Co-based Heusler compounds, with their great diversity, give us the opportunity to tune our compounds (e.g. the number of valence electrons, spin-orbit coupling strength, etc.) across different compositions in order to get the desired properties.


We perform ab-initio calculations based on the density functional theory (DFT)~\cite{Hohenberg-PR64,Kohn-PR65} and the generalized gradient approximation (GGA) for the exchange-correlation potential~\cite{PBE} (more details in Supplemetary Material).
The spin-polarized band structure and density of states (DOS) of ZrCo$_2$Sn are calculated within GGA+U without SOC, shown in Fig.~\ref{fig1}(c) and Fig.~\ref{fig1}(d) respectively. The value of U was chosen to be 3 eV, which provides Curie temperature close to the experimental value~\cite{TC} and reproduces the measured magnetic moment (see Fig. 5 of Supplementary Material section). The Curie temperature was computed by means of random-phase approximation (RPA) by S. V. Tyablikov~\cite{Tyablikov}. 
We have tested U with different functionals, such as LDA~\cite{PhysRevB.23.5048}, PBE~\cite{PBE} and PBEsol~\cite{PhysRevLett.80.891}, and get the same result (detailed calculation of U as a fucntion of $T_C$ and the magnetic coupling constants can be found in the Supplementary Material section). In the following we will use the PBE pseudopotential.

\begin{figure}
\centering
 \includegraphics[width=0.48\columnwidth]{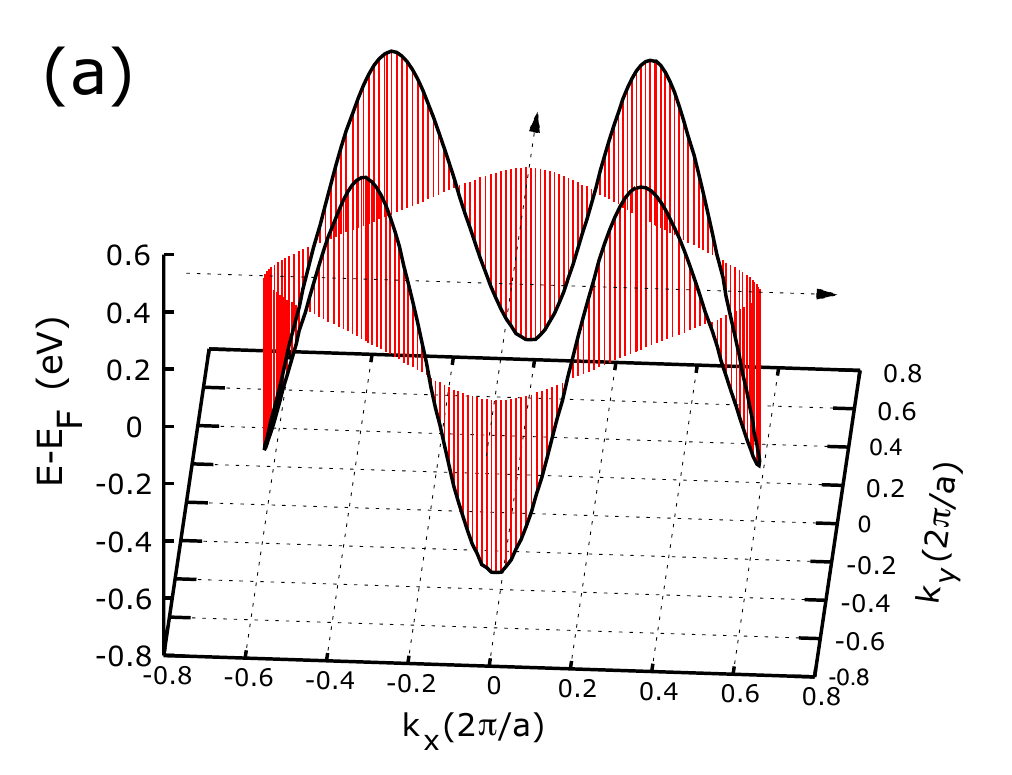}
 \includegraphics[width=0.48\columnwidth]{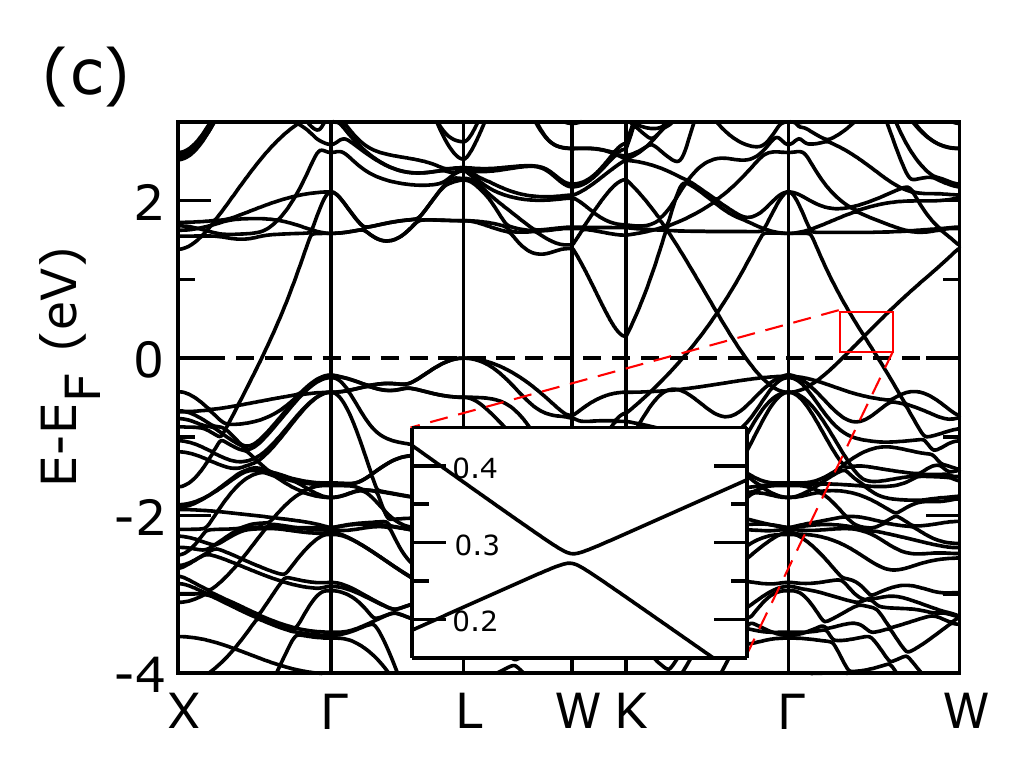}
 \includegraphics[width=0.48\columnwidth]{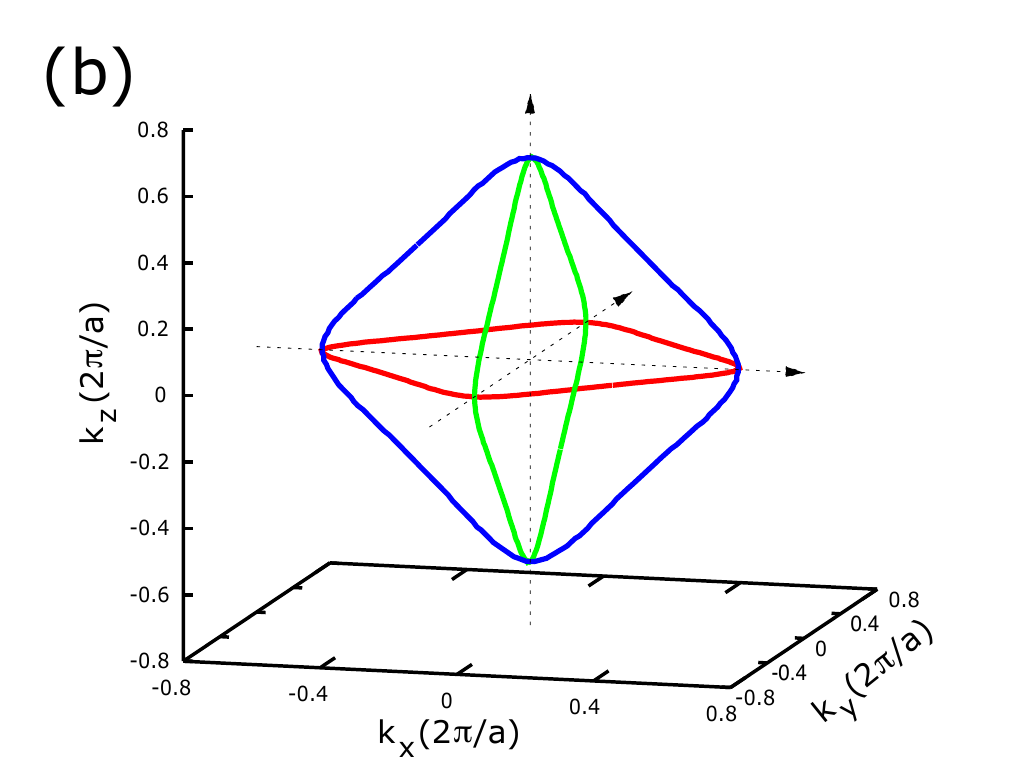}
 \includegraphics[width=0.48\columnwidth]{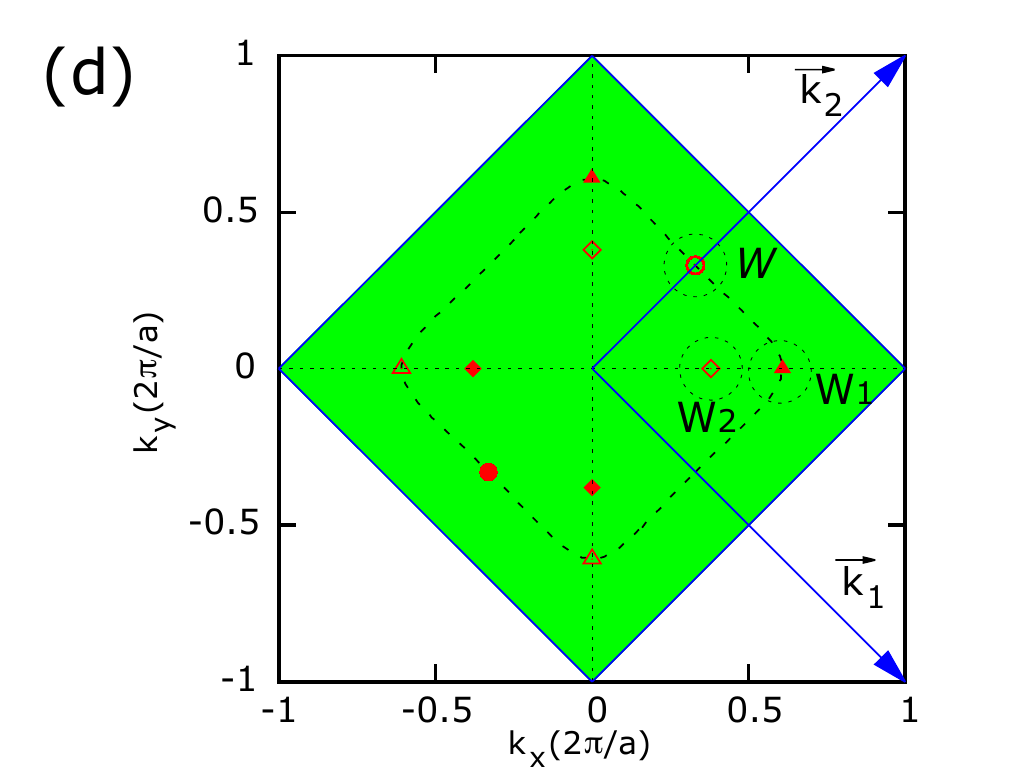}
    \caption{ (a) The nodal line in the $xy$ plane has a big dispersion. (b) Three nodal lines in three planes in three-dimensional (3D) {\em k}-space. (c) The SOC band sturcuture of ZrCo$_2$Sn with the [110] magnetism, opening a small gap in the $\Gamma$W direction in the inset. (d) Weyl points emerge with SOC. The independent $W$, W$_1$ and W$_2$ points are clearly indicated in the figure (top view) - the remaining ones can be obtained by symmetry. $W$ and W$_1$ is in the $xy$ plane, while the W$_2$ is out of the plane. The Chern numbers can be calculated with the Wilson-loop method applied on a sphere (illustrated as dashed circles) enclosing a Weyl point. The filled (unfilled) symbels indicate Chern number +1 ($-1$). Furthermore, the 001-surface lattice vectors are also given as ${\vec k}_1(\frac{2\pi}{a},-\frac{2\pi}{a})$ and $ {\vec k}_2(\frac{2\pi}{a},\frac{2\pi}{a})$, and the corresponding surface BZ is painted in green.}
    \label{fig2}
\end{figure}

In Fig.~\ref{fig1}(c), we see a band gap in the minority states around $E_F$, which has been double-checked with the modified Becke-Johnson exchange potential~\cite{mbj2009}. This suggests a good half-metallic property, which is consistent with the experimental investigation of the spin-resolved unoccupied density of states of the partner compound TiCo$_2$Sn~\cite{felser2009}. 
The calculated partial DOS, shown in Fig.~\ref{fig1}(d), suggests the states of ZrCo$_2$Sn around $E_F$ are mostly of d character, namely Co-d and Zr-d states.
After including SOC, the calculated band structure in Fig.~\ref{fig2}(c) shows the SOC has little influence on the electronic structure and the half-metallic ferromagnetism, because the SOC strengths of both Co and Zr are small.
We performed ab-initio calculations to determine the energetically most favorable magnetization direction - the  $[110]$ easy axis. However, this magnetic configuration is energetically very close to the [100]. 
we assume the magnetism is along the $[110]$ axis in the main text. The analysis of the $[100]$ magnetism is presented in Supplementary Material, which also shows topological properties such as Weyl points and nodal lines in the band structure.


In the spin-polarized calculation, neglecting SOC, the spin and the orbitals are independent and regarded as different subspaces. In that sense, the spatial crystal symmetries of $O_h^5$ have no effect on the spin degree of freedom, and the two spin channels are decoupled. Once SOC is considered, the two spin channels couple together, and  symmetries can decrease depending on the direction of the spontaneous magnetization. For instance, the mirror reflection $M_z$ is a symmetry without SOC included. With SOC it is broken as the magnetization is along $[110]$. However, the product between time-reversal and reflection symmetries ($TM_z$) is still a magnetic symmetry, even though SOC is included. The corresponding magnetic space group of [110] spin-polarization contains only 8 elements formed by three generators: inversion $\cal I$, 2-fold rotation $C2_{110}$ around 110 axis and the product ($TC2_z$) of time-reversal and rotation $C2_z$, which allows for the existence of Weyl points (WPs) in the $xy$-plane~\cite{soluyanov2015nature}.

We first elucidate the band topology in the absence of  SOC. From the spin-polarized  band structure in Fig.~\ref{fig1}(c), we can observe that two band crossings occur along $\Gamma$X, $\Gamma$W and $\Gamma$K in the majority spin. Actually, these crossings are in the $xy$-plane, which respects the mirror symmetry $M_z$. Bands within this plane can be classified by mirror eigenvalues $\pm$1. Further symmetry analysis shows that the two crossing bands belong to opposite mirror eigenvalues, giving rise to a nodal line in the $xy$ plane~\cite{PhysRevX.5.011029,hasanatcomm}.
The energy of the nodal line disperses dramatically in the $xy$ plane, as shown in Fig.~\ref{fig2}(a).  The minimum of this dispersion is in the $\Gamma$X (or [100]) direction and the maximum is in the $\Gamma$K (or [110]) direction. In addition to the nodal line in the $xy$-plane, two similar nodal lines are also found in the $xz$-plane and $yz$-plane related by a $C4$ rotation around the $x, y$ coordinate axis. As a result, the three nodal lines in different planes intersect at six different points as depicted in Fig.~\ref{fig2}(b).

Once introducing SOC, with a magnetization along the $[110]$ direction, the mirror symmetries $M_z$, $M_x$ and $M_y$ are all broken. In the absence of other symmetries, these nodal lines in the mirror planes would become fully gapped. However, along the magnetization $[110]$ direction, a pair of Weyl points survive, protected by the $C2_{110}$ rotation. Namely, the crossing bands belong to different $C2_{110}$ eigenvalues $\pm i$ on the high-symmetry line. The coordinates of these WPs ($W$), related by inversion $\cal I$, are given in Tab.~\ref{posi}. Their location and Chern numbers are illustrated in Fig.\ref{fig2}(d). An inversion eigenvalue argument shows us that we must have $4k+2$, $k \in {\mathcal{Z}}$, number of Weyls in this system (see Supplementary Material): the product of the inversion eigenvalues of the occupied bands at the inversion symmetric points is $-1$, signaling the presence of an odd number of pairs of Weyls \cite{PhysRevB.83.245132}. 

\begin{figure}
\centering
 \includegraphics[width=0.98\columnwidth]{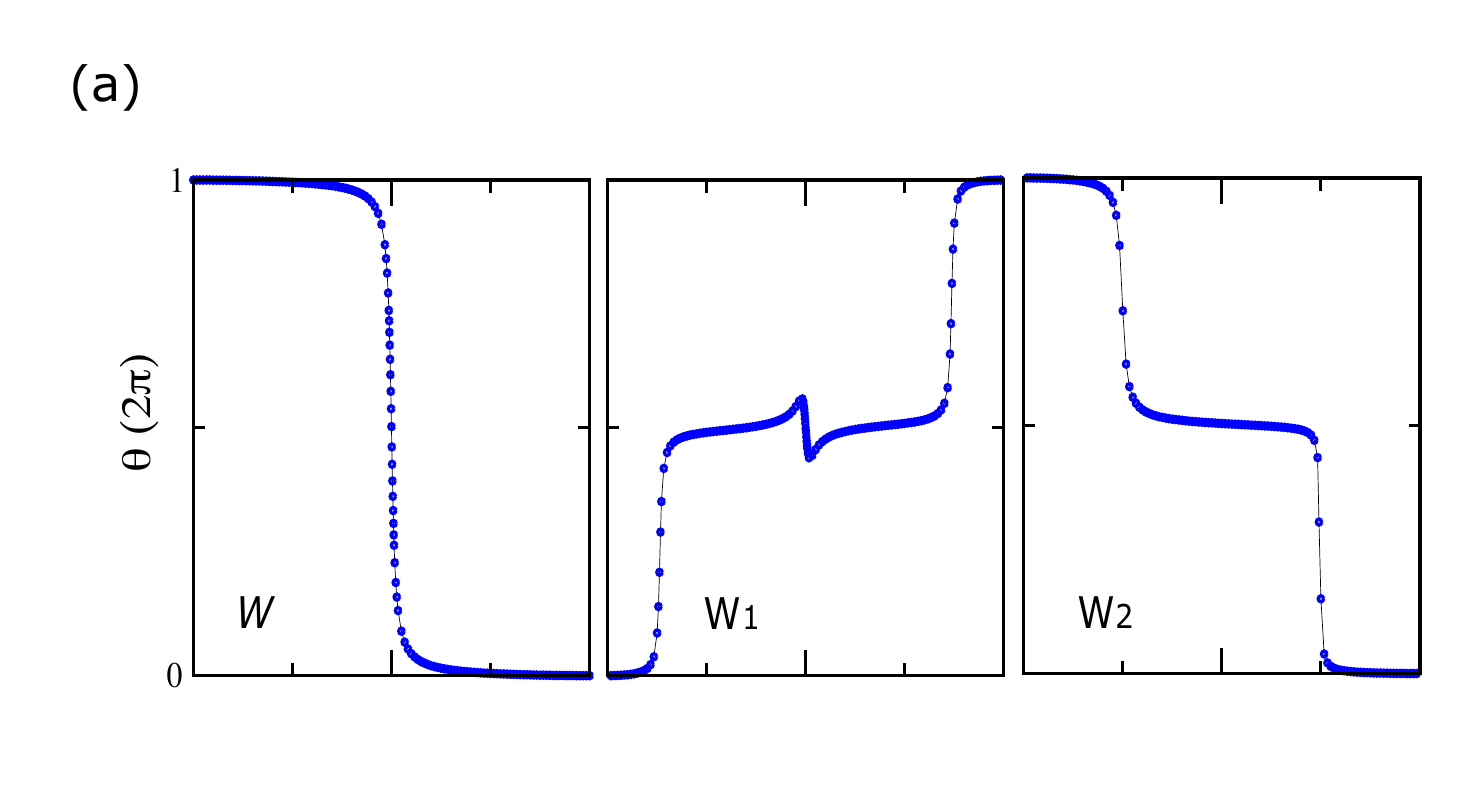}
 \includegraphics[width=0.98\columnwidth]{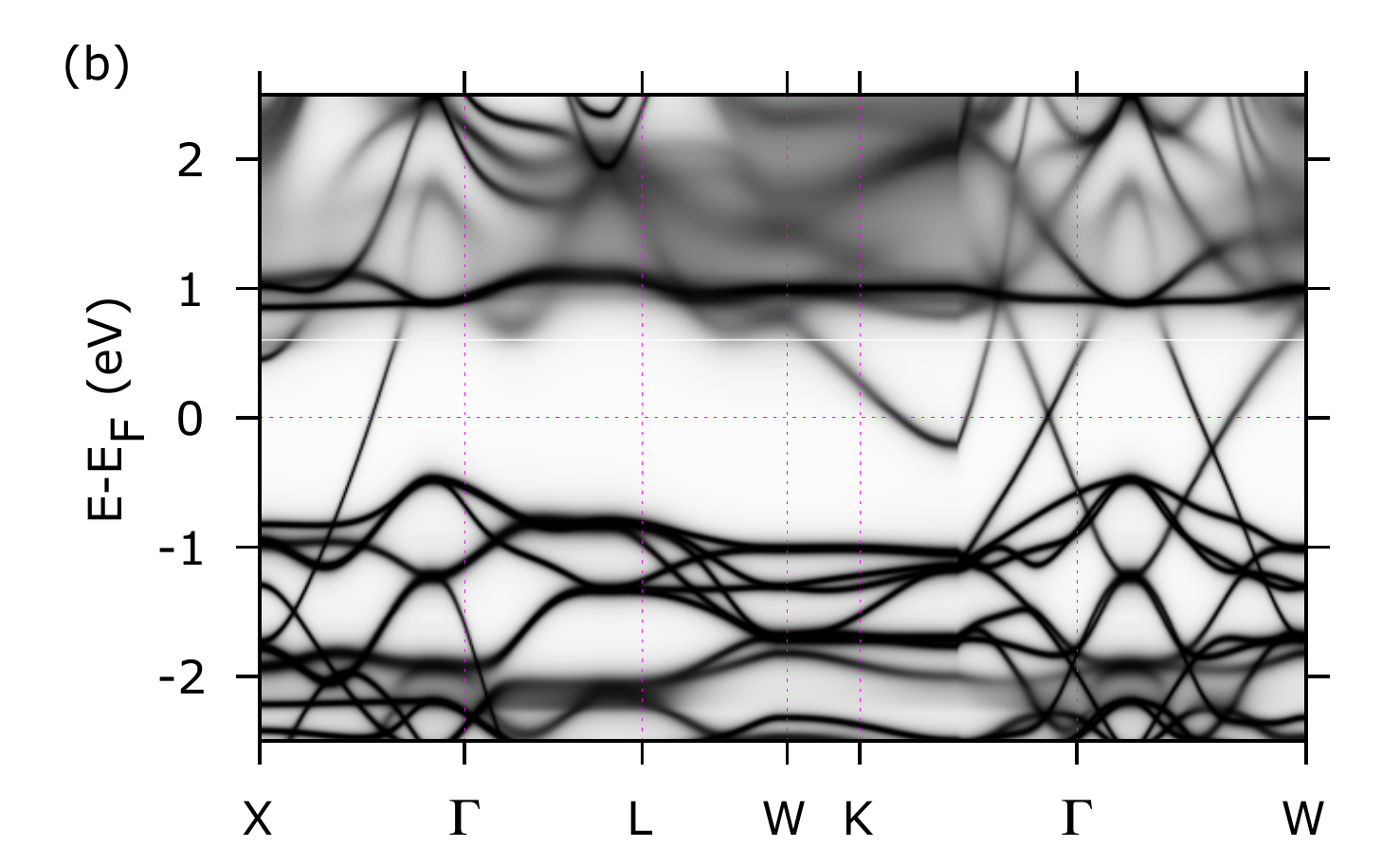}
    \caption{ (a) The flow chart of the average position of Wannier centers obtained by the Wilson-loop method applied on a sphere that encloses a node~\cite{soluyanov2015nature}. The average center shifts downwards, corresponding to Chern mumber $-1$ for $W$ and W$_2$, while it shifts upwards, suggesting the Chern mumber of W$_1$ is +1.  (b) Calculated Bloch spectral function of Zr$_{0.725}$Nb$_{0.275}$Co$_{2}$Sn along high symmetry lines.   }
     \label{fig3}
\end{figure}

In addition, deriving from the nodal line in the $xy$-plane, other four Weyls are found slightly away from the coordinate axis in the plane. The presence of Weyls in a high symmetry plane is allowed by the antiunitary symmetry $TC2_z$ \cite{soluyanov2015nature}. The quadruplet Weyls (W$_1$) are related to each other by $\cal I$ and $C2_{110}$. Their precise positions and topological charges are presented in Tab.~\ref{posi}. 
After carefully checking other two nodal lines, we see that a third kind of Weyl points (W$_2$), different from the previous two kinds, does not prefer any special direction, but distributes near the $xz$- and $yz$-plane. As these Weyls are generic points without any little-group symmetry, the octuplet Weyls W$_2$ are related by all the three generators of the magnetic group. As a result, the position of the W$_2$ changes considerably from TiCo$_2$Sn, to ZrCo$_2$Sn to HfCo$_2$Sn, following the nodal lines without SOC. In principle, this kind of Weyls are not stable (in contrast, the $W$-Weyls are stablized by the $C2_{110}$ symmetry on the [110] axis), as they can be moved close in the $z$-axis and thereby annihilate with each other. The average charge centers obtained by the Wilson-loop method on the spheres ($W$, W$_1$ and W$_2$) are presented in Fig.~\ref{fig3}(a). All the Chern numbers of the three Weyls are shown in Table~\ref{posi}, and their positions are shown in Fig.~\ref{fig2}(d). As the energy level of W$_1$ is very low and W$_2$ could be removed by tuning SOC, we will focus on the doublet Weyls $W$.

\begin{table}[h!]
\begin{center}
\caption{WPs of ZrCo$_2$Sn. The Weyl points' positions (in reduced coordinates $k_x$, $k_y$, $k_z$), Chern numbers, and the energy relative to the $E_{\mathrm{F}}$ of the unalloyed  compound. The WPs in ZrCo$_2$Sn are formed by two bands which in the absence of SOC would form  nodal lines. $W$ and W$_1$, are stable in the $xy$ plane, while the W$_2$ are stable out of the plane. The coordinates of the other Weyl points are related to the ones listed by the symmetries, $\cal I$, $C2_{110}$ and ${\cal T}C2_z$.}
\label{posi}
\begin{tabular}{cccc}
\hline
\hline
Weyl points & coordinates & Chern number & $E-E_{\mathrm{F}}$ \\
& ($k_x\frac{2\pi}{a},k_y\frac{2\pi}{a},k_z\frac{2\pi}{a}$) &  & (eV) \\
\hline
$W$ & $(0.334,0.334,0)$ &$-1$ &$+0.6$\\
W$_1$ & $(0.58,-0.0005,0)$ &$+1$ &$-0.6$ \\
\hline
W$_2$ & $(0.40,0.001,\pm 0.28)$ &$-1$ &$+0.3$ \\
\hline
\hline
\end{tabular}
\end{center}
\end{table}

We now focus on the two $W$ type Weyls (located at 0.6 eV over the Fermi level in the K$-\Gamma$ (or [110]) direction. Our goal now is to tune the energy of the WP to the Fermi level. For this purpose we consider other compounds with the same stoichiometry, more electrons and similar lattice parameter.As we mentioned before, NbCo$_2$Sn, which have the same crystal structure~\cite{nbco2sn1996}, contains one more electron per a unit cell than that of ZrCo$_2$Sn. 
Therefore, one can expect that alloying ZrCo$_2$Sn with Nb in the Zr site would shift down the WP energy while keeping the main band topology unchanged. 
\begin{figure}
\centering
\includegraphics[width=0.98\columnwidth]{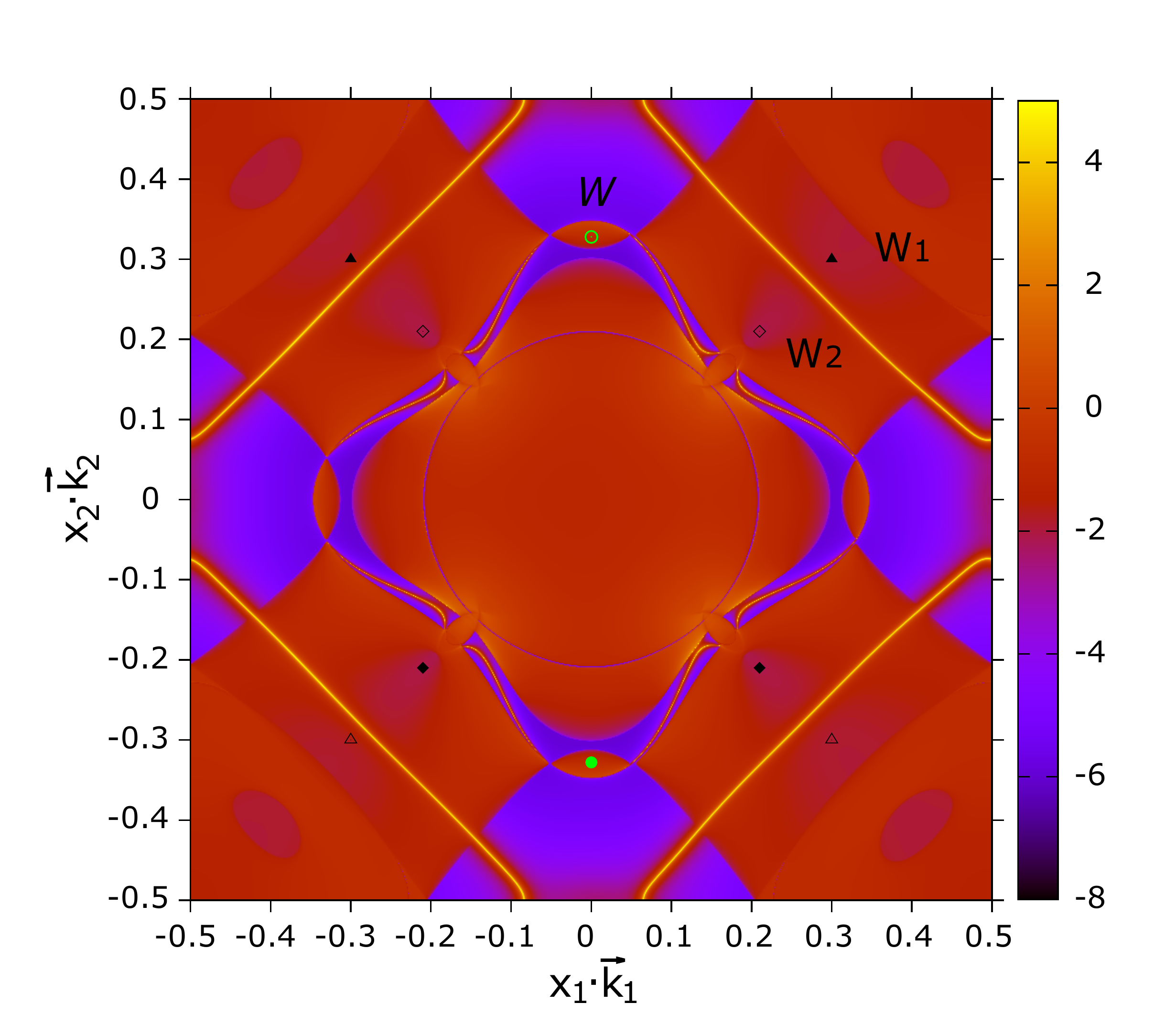}
    \caption{Bloch spectral function of the (001)-surface at 0.5 eV above the Fermi level for ZrCo$_2$Sn. In the (001)-surface Brillouin zone, the surface {\bf k}-points are represented by $x_1 \vec k_1+x_2\vec k_2$. The surface lattice vectors ($\vec k_1$ and $\vec k_2$) are illustrated, and the corresponding surface BZ is shown as green-colored area in Fig.~\ref{fig2}(d) (notice the $\pi/2$ rotation of the surface BZ). Only the bulk projections of $W$ (yellow-colored) are separated from the Fermi surfaces projections.  The bulk projections of W$_1$ and W$_2$ (black-colored) sit inside the projection of the bulk Fermi surfaces. The Fermi arcs connecting to the bulk projections of $W$ are shown. The large, square surface states is of a trivial nature. The color code represents $log(\rho)$.}
     \label{fig4}
\end{figure}  
Using a first-principles Green's function method, we dope ZrCo$_2$Sn with Nb. Disorder effects were taken into account within a coherent potential approximation (CPA)~\cite{PhysRevB.5.2382}. Varying Nb content, we search for a concentration, which brings the $W$ Weyls to the Fermi level.
Fig.~\ref{fig3}(b) shows the calculated spectral function for Zr$_{1-x}$Nb$_{x}$Co$_{2}$Sn (with $x=$ 0.275). By inversion ($\cal I$) symmetry, there exist another Weyl point separated in k-space by $\sim$$2\pi$ with the same energy. In the same line, the experimental existence of VCo$_2$Sn \cite{nbco2sn1996} also motivate us to dope the partner compound Ti$_{1-x}$V$_x$Co$_2$Sn as well, and our calculations suggest $x= 0.1$.

Given that the Weyl nodes $W$, W$_1$ and W$_2$ all resulted from the connected nodal lines in the absence of SOC, a large residual Fermi surface has a projection on any surface of the material. Hence the Fermi arcs emanating from the $W$ Weyl points are interrupted by the residual projection of bulk Fermi surfaces on the surface of the material. However, the Fermi arc signatures of $W$-Weyls are still clear as can be seen in  Fig.~\ref{fig4} where we plot the surface spectral function for the (001)-surface of ZrCo$_2$Sn. Since the bulk Fermi surface projections where W$_1$ and W$_2$ are located overlap, the Fermi arcs connections are not guaranteed at some certain energies. Furthermore, a trivial square surface state is found as well, due to the hanging bonding at the surface.

In conclusion we have predicted theoretically that a new family of Co-based magnetic Heuslers realize Weyl systems with several Weyl nodes whose position in energy can be tuned by alloying. We have performed ab-initio calculations of a representative ferromagnetic compound ZrCo$_2$Sn. For the $[110]$ magnetization we find two Weyl points related by $\cal I$ symmetry situated on the same axis. By means of a first-principles Green's function method, we doped the ZrCo$_2$Sn with Nb and showed that these two Weyls can be shifted to the Fermi level. Finally, we obtained the Fermi arc structure of this class of materials. 
This discovery shows a way to the realizing the hydrogen atom of Weyl materials and provides a promising platform for studying exotic properties of magnetic Weyls in experiments.

 \textbf{Acknowledgments} We thank Zhida Song for providing the overlap calculating code. Z. Wang and B.A. Bernevig were supported by Department of Energy DE-SC0016239, NSF EAGER Award NOA-AWD1004957, Simons Investigator Award, ONR-N00014-14-1-0330, ARO MURI W911NF-12-1-0461, NSF-MRSEC DMR-1420541, Packard Foundation, Schmidt Fund for Innovative Research and the National Natural Science Foundation of China (No. 11504117). MGV and EVC acknowledge partial support from the Basque Country Government, Departamento de Educaci\'{o}n, Universidades e Investigaci\'{o}n (Grant No. IT-756-13), the Spanish Ministerio de Econom\'{i}a e Innovaci\'{o}n (Grant No. FIS2010-19609-C02-01 and FIS2013-48286-C2-1-P), the FEDER funding and the Saint Petersburg State University (Project No. 15.61.202.2015). AE aknowledges the Joint Initiative for Research and Innovation within the Fraunhofer and Max Planck cooperation program and the DFG Priority Program 1666 ``Topological Insulators". RJC and NPO were supported by MURI award for topological insulators (ARO W911NF-12-1-0461).\\

\bibliography{MagneticHeusslers}



\section*{SUPPLEMENTARY MATERIAL}

\appendix

\section{Co-based family of Heusler compounds}
\begin{figure}[t!]
\centering
\includegraphics[width=0.98\columnwidth]{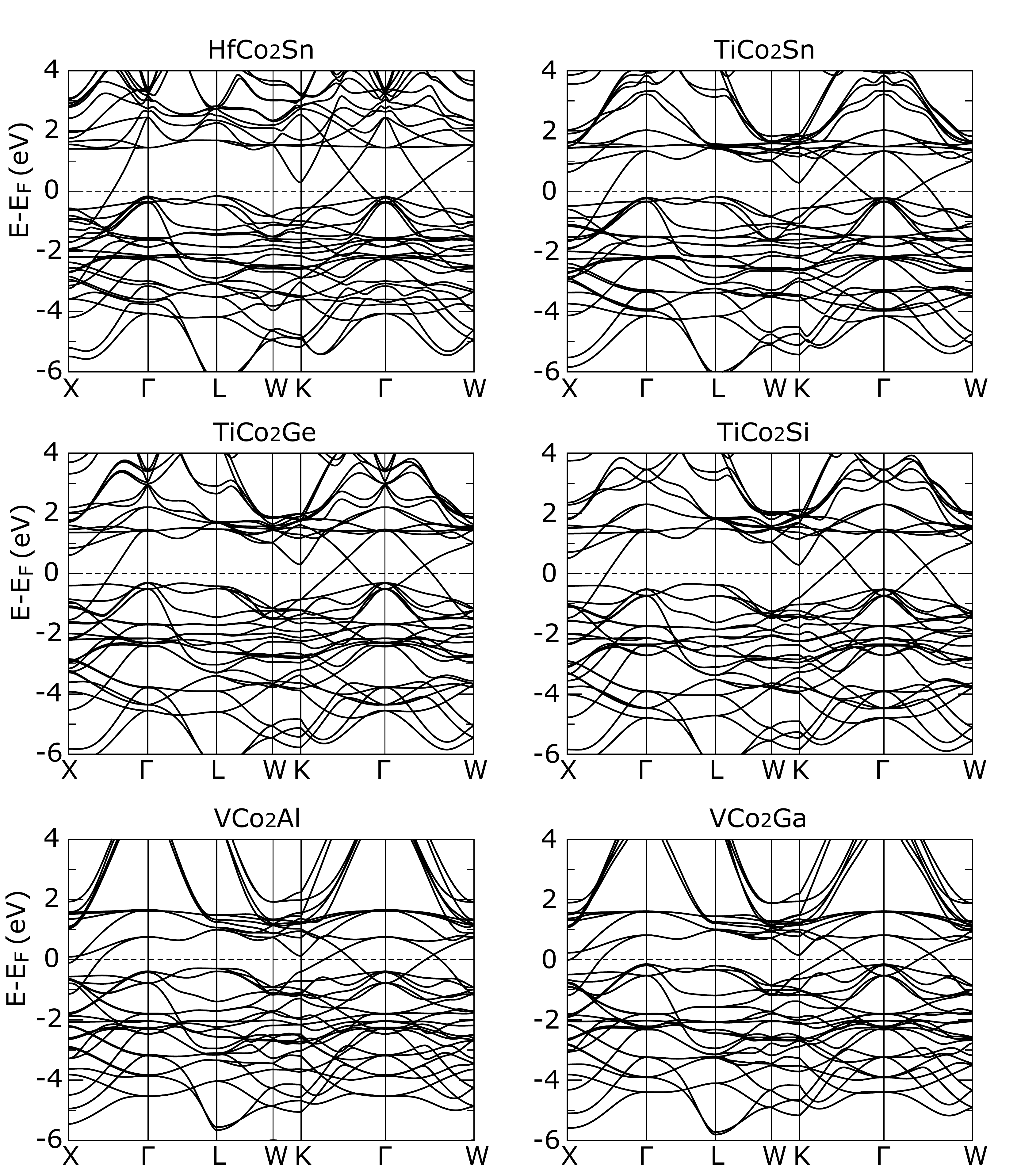}
    \caption{ Band structure of Co-based family compounds along high symmetry lines.}
     \label{app1}
\end{figure}

In this study we present a Co-based magnetic Heusler materials based family: HfCo$_2$Sn, ZrCo$_2$Sn,  TiCo$_2$Sn, TiCo$_2$Ge, TiCo$_2$Si, VCo$_2$Al and VCo$_2$Ga, their lattice parameters and references are shown in Table \ref{cos}. The band structure of these compounds is depicted in Fig.\ref{app1}
\begin{table}[h!]
\begin{center}
\begin{tabular}{ccc}
\hline
\hline
Compound & Lattice parameter & Reference \\
HfCo$_2$Sn & 6.220 \AA & \cite{HfCo2Sn} \\
TiCo$_2$Sn & 6.067 \AA & \cite{TiCo2Sn} \\
TiCo$_2$Ge &  5.831 \AA & \cite{TiCo2Ge} \\
TiCo$_2$Si & 5.743 \AA & \cite{TiCo2Si} \\
VCo$_2$Al & 5.772 \AA & \cite{TiCo2Si} \\
VCo$_2$Ga & 5.779 \AA & \cite{VCo2Ga}\\
\hline
\hline
\end{tabular}
\caption{Lattice parameters and references of the Co-based family compounds.}
\label{cos}
\end{center}
\end{table}

\section{Calculation Methods}

The first-principles calculations have been performed within the DFT as implemented in the Vienna Ab-initio Simulation Package (VASP) \cite{Kresse199615,PhysRevB.48.13115}.  The interaction between ion cores and valence electrons was treated by the projector augmented wave method \cite{vaspPaw}.
The orbitals considered as valence electrons in the atomic pseudo-potential for the representative compound ZrCo$_2$Sn were: Zr (4$s^2$4$p^6$5$s^2$4$d^2$),  Co (3$d^8$4$s^1$) and Sn (5$s^2$5$p^2$). 
To describe the exchange-correlation energy, we used the General Gradient Approximation (GGA) with the Perdew-Burke-Ernkzerhof (PBE) parameterization \cite{PhysRevLett.80.891}. In order to account for the magnetic character of the system we performed spin polarized calculations, with the full crystal symmetry implemented per spin species. 
The Hamiltonian contains  the scalar relativistic corrections and the spin-orbit coupling is taken into account by the
second variation method \cite{PhysRevB.62.11556}. Since the $d$-electrons of Cobalt are strongly correlated we introduce on-site Coulomb repulsion \cite{PhysRevB.57.1505} of $U$ $=$ 3 eV, a value chosen to  reproduce the $T_C$ temperature of the system \cite{ZrCurie}.
We use the experimental lattice parameters and atomic positions in the unit cell. A \emph{k}-point grid of (20$\times$20$\times$20) for reciprocal space integration and 400 eV energy cut-off of the plane wave expansion have been used to get a residual error on atomic forces below 1 meV/\r{A} and a fully converged electronic structure including spin-orbit coupling (SOC).

The electronic structure calculations of alloys were performed using a self-consistent full-charge-density Green's function method \cite{KORRINGA1947392,PhysRev.94.1111,0953-8984-13-38-305} in a scalar relativistic scheme. We used 24 Gaussian quadrature points to carry out a complex energy contour integration, while for the integration over the Brillouin zone we used a mesh of (20$\times$20$\times$20) k points. Substitutional disorder was treated within the coherent potential approach (CPA)~\cite{PhysRevB.5.2382}.

\section{Magnetic properties}

\begin{figure}[h!]
\centering
\includegraphics[width=0.8\columnwidth]{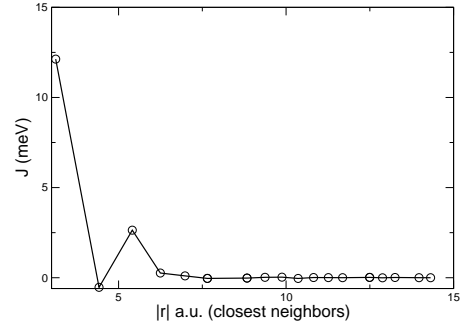}
    \caption{ Calculated exchange interaction between Co atoms for ZrCo$_2$Sn. The $x$ axis corresponds to the effective distance between closest neighbors.}
     \label{js}
\end{figure}

In this section we present the magnetic properties and a detailed calculation of U for our representative compound ZrCo$_2$Sn. The exchange constants of the Heisenberg Hamiltonian were obtained using the magnetic force theorem as it is implemented within multiple scattering-theory~\cite{Lich}. Our calculations, as shown in Fig.~\ref{js}, reveal that our system is coupled by short ferromagnetic interaction.

\begin{figure}[t!]
\centering
\includegraphics[width=0.8\columnwidth]{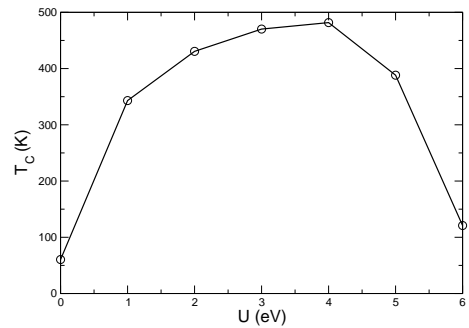}
    \caption{ Calculations of the Curie Temperature as a function of U.}
     \label{UTC}
\end{figure}

Fig.~\ref{UTC} shows the calculation of $T_C$ using RPA as a function of U. As we can observe only the value of U $=3$eV reproduces the experimental value of  $T_C=448$K~\cite{nbco2sn1996}.

\section{Experimental realization}

\begin{figure}[h!]
\centering
    \includegraphics[width=0.98\columnwidth]{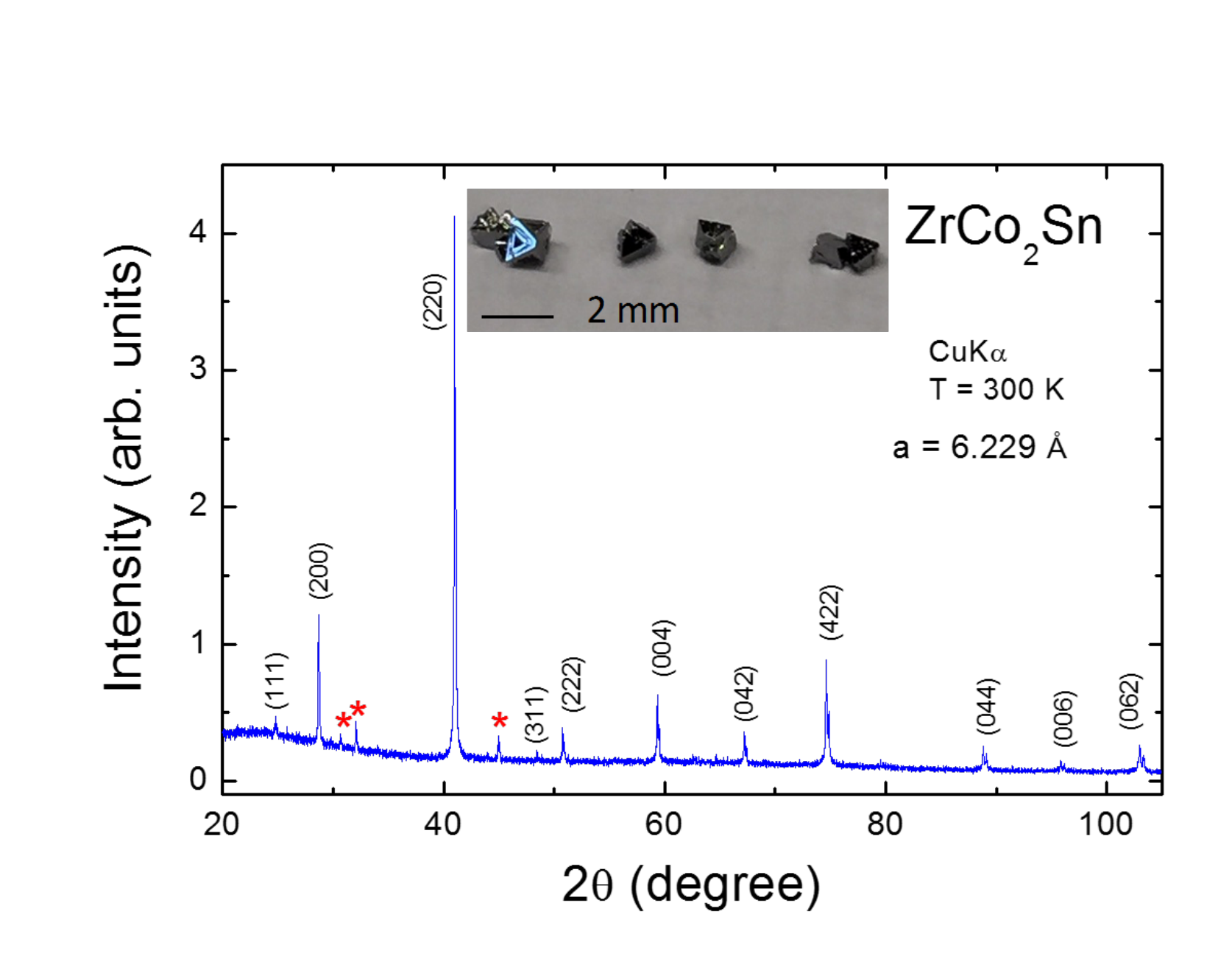}
    \caption{The X-ray diffraction pattern recorded for a powder specimen of crushed ZrCo$_2$Sn crystals. The peaks marked by red asterisks are due to the Sn-flux.}
     \label{Exp1}
\end{figure}

\begin{figure}[h!]
\centering
    \includegraphics[width=0.98\columnwidth]{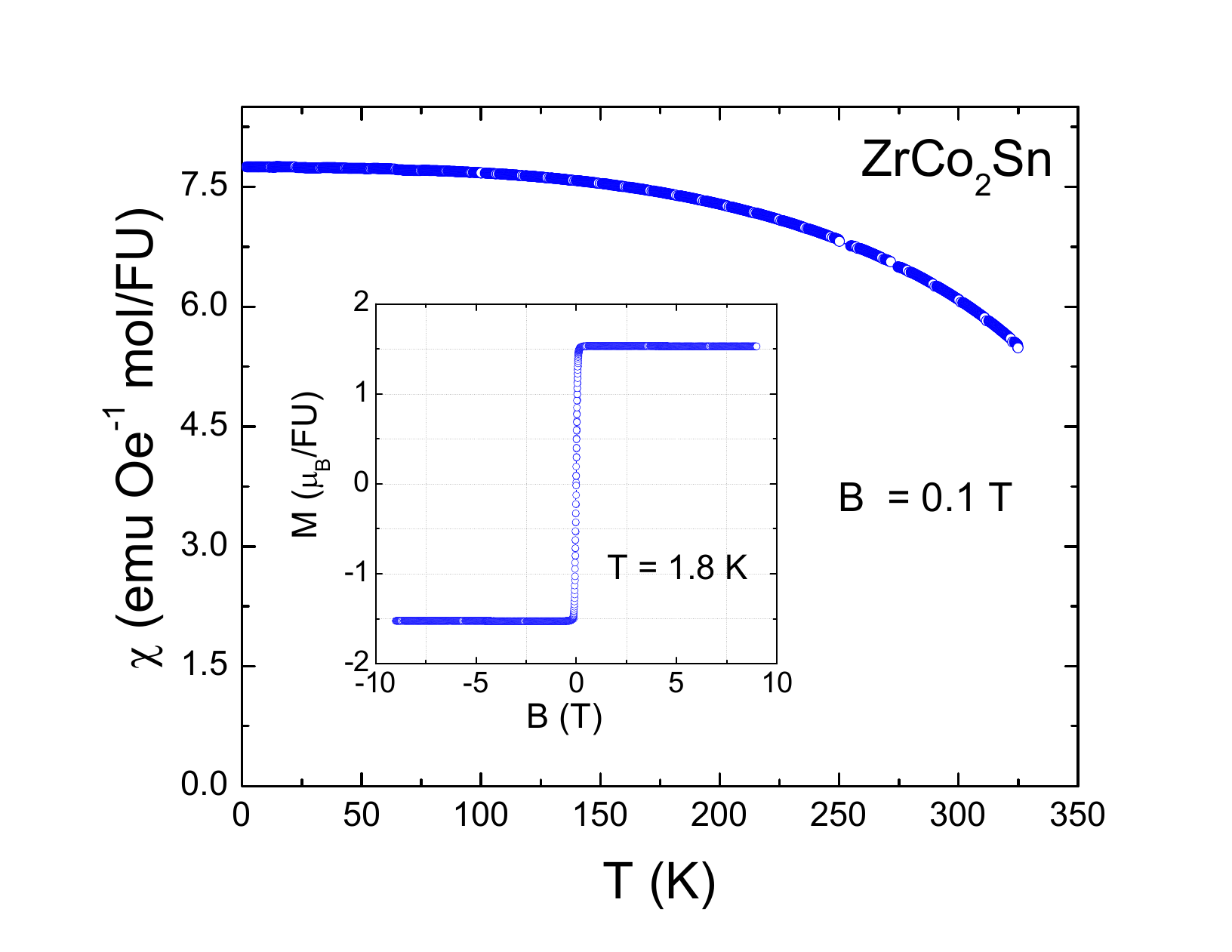}
    \caption{The main panel shows the temperature dependent magnetic susceptibility, $\chi$ = M/H, at H = 1000 Oe, for a single crystal of ZrCo$_2$Sn. The nonzero, high $\chi$ values in the entire temperature range indicate that the as-grown crystals are ferromagnetic in nature, with a $T_C$ well above the room temperature. In the inset is a plot of M vs. H for a single crystal recorded at 1.8 K. The estimated magnetic moment ~1.52  $\mu$B/formula unit, and the soft magnetic character is shown.}
     \label{Exp2}
\end{figure}

The high quality ZrCo$_2$Sn single crystals of mm size were grown by the Sn-flux method. The Zr, Co and Sn elements were placed in a 3:25:75 atomic ratio in an alumina crucible and then sealed under vacuum in a quartz glass tube. The ampoule was heated to 1150 $^{\circ}$C, where it was held for 6 hours, and then cooled at 2 $^{\circ}$C/hr. to 1000 $^{\circ}$C, when it was centrifuged to extract the crystals from the flux. The crystals grew with well-developed cube-truncated-octahedron morphologies, as shown in the inset of Fig.~\ref{Exp1}. The Heusler phase for the crystals was determined by recording powder X-ray diffraction patterns (CuK$\alpha$ radiation) on crushed and ground crystals. The main panel in Fig.~\ref{Exp1} shows an X-ray diffraction pattern recorded on a powdered sample, in which the match is to the reported Heusler phase pattern given in the ICSD. The lattice parameter of the crystals is found to be a = 6.229 $\AA$. The diffraction peaks are quite sharp and the K$\alpha$1 and K$\alpha$2 splitting of the peaks can be seen clearly for higher angle peaks, an indication of the crystal quality. This shows that the grown crystals are of high quality. The temperature and field dependent magnetizations for several single crystals were measured in a Quantum Design PPMS system. Representative data are shown in Fig.~\ref{Exp2}. From the saturation magnetization plots the magnetic moment is found to be ~ 1.52 $\mu$B/formula unit, which is close to the earlier reported experimental values~\cite{Exp1,Co2ZrSnAHE}.

\section{Inversion Symmetry Characterization}

With broken time-reversal, when inversion symmetry is present, a simple diagnostic can be applied to find out if an odd number of pairs of Weyl nodes ($2(2n+1)$ Weyl Fermions) are present in a material \cite{PhysRevB.83.245132}. Take the product of the inversion eigenvalue $\zeta_n({\bf K}_i)$ of all bands $n$  below the Fermi level, over all bands $n$ at all inversion symmetric points ${\bf K}_i$ (which are also time-reversal invariant momenta):
\beq
\prod_{{\bf K}_i = - {\bf K}_i} \prod_{ E_n ({\bf K}_i)< E_f} \zeta_{n}({\bf K}_i) 
\eneq If the product is $-1$ then an ODD number of pairs of Weyls exists in the bulk, which means that Weyl nodes \emph{must} exist. If the product is $+1$ then it is not possible (with inversion symmetry only) to easily deduce if a nonzero number of pairs or a zero number of pairs is present in the system. For our system, we find the inversion eigenvalues in Table \ref{pts} with product $-1$, and hence an odd number of Weyl pairs needs to be present in our system.

\begin{table}[h!]
\begin{center}
\caption{ Parities at three independent time-reversal invariant momenta (TRIM). At each TRIM, the first row is for the majority channel, while the second one is for the minority channel. ``;" refers to the Fermi level.}
\label{pts}
\begin{tabular}{cccccccccccccccc}
\hline
\hline
$\Gamma$ &+ &+ &$-$ &$-$ &$-$ &+ &+ &+ &$-$ &$-$ &+ &+ &+ &;&+ \\
         &  &  &$-$ &$-$ &$-$ &+ &+ &+ &+ &+ &+ &+ &+ &;&$-$ \\
\hline                                           
 $X$     &$-$ &+ &$-$ &+ &$-$ &$-$ &+ &+ &+ &$-$ &$-$ &$-$ &$-$ &;& + \\
         &  &  &$-$ &+ &+ &$-$ &$-$ &+ &+ &$-$ &$-$ &$-$ & $-$&;&+ \\
\hline                                           
 $L$     &$-$ &$-$ &$-$ &+ &+ &$-$ &$-$ &$-$ &+ &+ &+ &$-$ &$-$ &;&$-$ \\
         &  &  &$-$ &$-$ &$-$ &$-$ &$-$ &$-$ &+ &+ &+ &+ &+ &;&$-$ \\
\hline
$Product$ &  &  &  &  &  &  &  &  &  &  &  &  &  &$-$&  \\
              &  &  &  &  &  &  &  &  &  &  &  &  &  &+&  \\
\hline
\hline
\end{tabular}
\end{center}
\end{table}
%
 
%

\section{Symmetries of $2$-band Hamiltonian with $[110]$ magnetization}

Including SOC, we derive the little-group and the symmetries for potential Weyl nodes when the ferromagnetic magnetization if parallel to the $[110]$ direction. In this case, the following symmetry group remains: $P$; $C2_{110}$; $C2_z \cdot T$; $C2_{1\bar{1}0}\cdot T$. We now analyze the little group on different high-symmetry planes and axes to find out if  Weyls can appear generically on those planes and axes.

\subsection{Little group of the  $k$-space Hamiltonian on the $[110]$ axis}

The little group is $C2_{110}$, and bands with different eigenevalues under this symmetry can cross. Away from this axis, the crossing splits  and Weyl nodes \emph{can} exist on this axis.

\subsection{Little group of the $k$-space Hamiltonian on the $100$ (or equivalent) axis}

The only element of the symmetry group that maps $k_x, 0,0$ to itself is $C2_z \cdot T $ with matrix representation $C2_z \cdot T = i \sigma_x K $ with $K$ complex conjugation. The Hamiltonian on this axis is $H(k_x, 0,0)= d_i(k_x, 0,0) \sigma_i$ with the constraint:
\beq
[H(k_x, 0,0), i \sigma_x K ] =0 
\eneq which gives:
\beq
d_z(k_x,0,0)=0
\eneq Hence
\beq
H(k_x, 0,0) = d_x(k_x, 0,0) \sigma_x + d_y(k_x, 0,0) \sigma_y
\eneq
The Hamiltonian contains two parameters, one momentum, ($k_x$) giving rise generically to avoided crossings. No Weyls are possible  this line generically.

\subsection{Little Group of the $k$-space Hamiltonian on the $x y$ plane}

The little group of the $k_z=0, \pi $ planes is $C2_z \cdot T$, just as in the previous subsection. Going through an identical calculation, now the Hamiltonian is
\beq
H(k_x, k_y,0) = d_x(k_x, k_y,0) \sigma_x + d_y(k_x, k_y,0) \sigma_y
\eneq
The Hamiltonian contains two parameters, two momenta, and Weyl nodes \emph{can} generically live on this high symmetry plane.

\subsection{Little group of the $k$-space Hamiltonian on the $([1\bar{1} 0], [001])$ plane}

The little group is $C2_{110}\cdot P$. This is essentially the mirror group with mirror plane (spanned by $[1\bar{1} 0]$ and $[001]$) , which is a little group for momenta in that plane.  From
\beq
[H(k_x,k_y, k_z \in \text{plane}\quad1\bar{1} 0, 001), C2_{110}\cdot P]=0
\eneq
we find $d_z=0$, $d_x= d_y$ in this plane. Hence in this plane you have only one independent coefficient $d_x$ but $2$ momenta, so this plane can support \emph{nodal lines}.

\section{Symmetries and Weyl points\\ with $[100]$ magnetization}
Our ab-initio calculations suggest the [100]-magnetic configuration is energetically very close to the [110]. Therefore, we also preform the calculations for this magnetic configuration.  
When the ferromagnetic magnetization parallel to the [100] direction, the following symmetry group remains: $P$; $C4_x$; $C2_x\cdot P$; $C2_y\cdot T$; and  $C2_z\cdot T$. On the $x$-axis, one kind of Weyl points (the position of P$_1$ is given in Table \ref{posi100}) is protected by the rotation $C4_x$, which possesses Chern number $+2$. The nodal line in the $yz$-plane remains with SOC due to the mirror symmetry $C2_x\cdot P$. Deriving from the other nodal lines, another kind of WPs (P$_2$ in Table \ref{posi100}) is also found in the $xy$-plane ($xz$-plane), which repects $C2_z\cdot T$ ($C2_y\cdot T$), allowing for the existence of Weyl points in the plane~\cite{soluyanov2015nature}. 

\begin{table}[h!]
\begin{center}
\caption{WPs of ZrCo$_2$Sn. The Weyl nodes' positions (in reduced coordinates $k_x$, $k_y$, $k_z$), Chern numbers, and the energy relative to the $E_{\mathrm{F}}$. P$_1$ exist on the $x$-axis, and P$_2$-Weyls are stable in the $xy$- and $xz$-plane. The coordinates of the other Weyl points are related by $C4_x$ and $P$. The WPs related by $C4_x$ possess the same chirality, while ones related by $P$ possess the opposite chirality.}
\label{posi100}
\begin{tabular}{cccc}
\hline
\hline
Weyl points & coordinates & Chern number & $E-E_{\mathrm{F}}$ \\
& ($k_x\frac{2\pi}{a},k_y\frac{2\pi}{a},k_z\frac{2\pi}{a}$) &  & (eV) \\
\hline
P$_1$ & $(0.58, 0.0 ,0.0)$ &$+2$ &$-0.60$ \\
P$_2$ & $(0.36, 0.30,0.0)$ &$-1$ &$+0.55$ \\
\hline
\hline
\end{tabular}
\end{center}
\end{table}

\vspace{1cm}

\section{Additional Remarks}

Vesta\cite{Momma:db5098}, gnuplot\cite{Gnuplot_4.4}, pyProcar\cite{pyprocar} and Mayavi\cite{ramachandran2011mayavi} software packages were used to create some of the illustrations.

\end{document}